\documentstyle[pre,aps]{revtex}

\input psfig

\def\ev{{\bf e}}

\def\cv{{\bf c}}
\def\rv{{\bf r}}
\def\tv{{\bf t}}
\def\pv{{\bf p}}
\def\nv{{\bf n}}
\def\Ev{{\bf E}}
\def\Bv{{\bf B}}
\def\Av{{\bf A}}

\begin{document}
\draft
\title{Topological Inclusions in 2D Smectic-C Films}
\author{ by David Pettey, T.C. Lubensky and Darren Link $^{*}$}
\address{Department of Physics and Astronomy, University of Pennsylvania,
Philadelphia, PA 19104, USA \\
 $^{*}$ Condensed Matter Laboratory, Department of
Physics, University of Colorado, Boulder, CO 80309, USA }
\maketitle

\vspace{1.0in}
\begin{center}
Corresponding Author: David Pettey \\
Department of Physics and Astronomy \\
University of Pennsylvania \\
Philadelphia, PA 19104
\end{center}


\newpage
\begin{abstract}
In thin films of smectic-C liquid crystals, localized regions
containing additional smectic layers form circular inclusions that
carry a topological charge. Such inclusions nucleate a companion
topological defect.  These inclusion-defect pairs are modeled as
topological dipoles within the context of a one-coupling constant
approximation to the $2D$ Frank free energy.  Deviations of the dipole
direction from a preferred orientation cause the dipoles to acquire a
logarithmic charge.  Thermal fluctuations of the dipole direction are
calculated and found to be large, scaling as the logarithm of the
system size.  In addition to dipole-dipole interactions arising from
the topological charges, we also find that the thermal fluctuations of
the dipole directions are coupled through a preference for global
charge neutrality of the logarithmic charges.
\end{abstract}

\vskip2pc
\twocolumn
\narrowtext

\newpage
\section{Introduction}
\par 
Phases of matter with broken continuous symmetries exhibit topological
defects, which can be thermally excited or induced via external
stresses.  The surface interactions of a foreign inclusion within an
ordered medium can induce nontrivial distortions in the bulk medium
and lead to the nucleation of defect structures
\cite{Poulin97-1,Lubensky97-1,Poulin97-2,Meyer72-1}.  Even in cases
where no defects are present, the distortions that arise can often be
described by virtual defects within the foreign inclusion itself in
much the same way that electric fields outside a conductor can be
described by image charges within the conductor.  Two such
experimental systems which have recently been studied are Langmuir
monolayers \cite{Fang97-1} and nematic emulsions
\cite{Poulin97-1,Poulin97-2}.
\par
At sufficiently high density, a Langmuir monolayer forms an
orientationally ordered liquid-condensed phase in which the order is
characterized by a $2D$ unit-vector order parameter, which we call the
$\cv$-director \cite{Degennes,Chandra,Chandra86-1}.  Under appropriate
conditions a liquid-condensed region can contain isolated domains of
the rotationally isotropic liquid-expanded phase, that is, regions
wherein the orientational order is absent.  Along the boundary
separating the two phases, there is typically a preferred orientation
of $\cv$ relative to the tangent vector to the boundary.  The
inclusions (liquid-expanded regions) can thus lead to a distortion of
the order-parameter field of the ordered medium (liquid-condensed
region) surrounding them.
\par
Nematic emulsions are dispersions of droplets of nematic liquid
crystal in an isotropic host \cite{Drzaic}.  Inverse nematic emulsions
consist of surfactant coated spherical water droplets dispersed in a
nematic host \cite{Poulin97-1,Lubensky97-1,Poulin97-2}.  The
surfactant can be prepared such that the nematic director prefers to
align either parallel or perpendicular to a surfactant coated surface.
Thus the droplets distort the local nematic field and, in the case of
normal boundary conditions, the droplets themselves carry a
topological hedgehog charge which must be compensated for by creating
additional defect structures within the ordered medium.
\par
Experimental observations \cite{Poulin97-1,Poulin97-2,Fang97-1} in
both Langmuir monolayers and inverse nematic emulsions have been
explained theoretically with some success
\cite{Lubensky97-1,Fang97-1}.  Neither, however, is the simplest
system one could imagine for the study of the effects of foreign
inclusions in an ordered medium.  In Langmuir films the line tension
of the interface between the two phases is so low that orientational
order in the surrounding medium and the shape of the inclusion are
strongly coupled, greatly complicating the theoretical treatment.  In
nematic emulsions, the elastic theory describing director distortions
is highly non-linear, and as a result, it is very difficult to find
exact solutions for the director around a droplet-defect pair.  In
this paper, we report observations of freely suspended inclusions in
another ordered system, smectic-C films, and we develop a theory for
their energetics.
\par
\begin{figure}
\psfig{figure=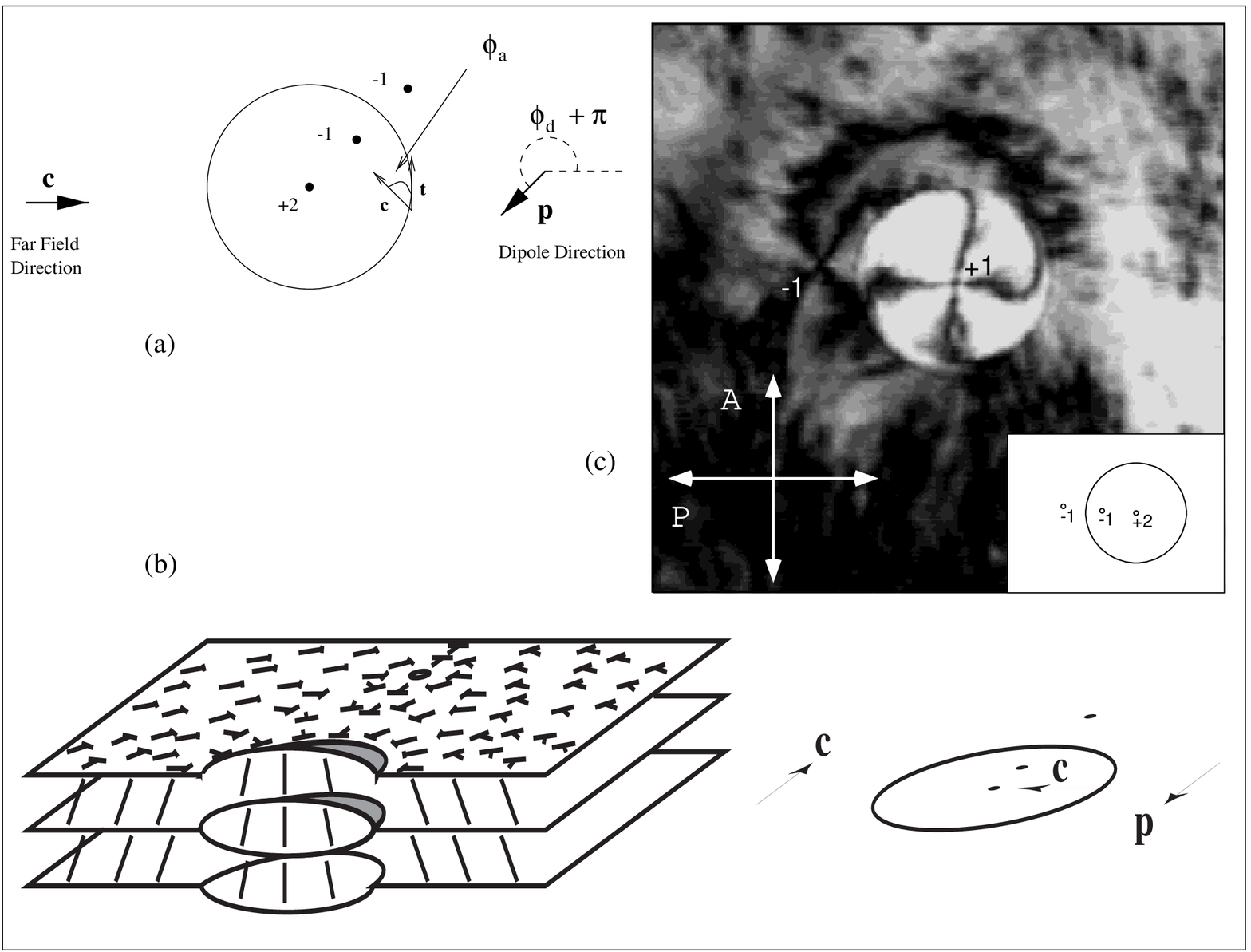,width=16.0cm}
\caption{
        }
\label{setup}
\end{figure}

There are several types of circular inclusions with fixed boundary
conditions that can be inserted into a freely suspended smectic-C
film.  The type we are concerned with here are islands, places where
the film contains an integral number of additional smectic layers (Fig
\ref{setup}).  Islands can be introduced to a freely suspended film by
(1) rapidly reducing the film area, (2) blowing across the film to
pull material from the meniscus region or (3) heating the film near
the smectic-to-isotropic or smectic-to-nematic phase transition, at
which point isotropic or nematic droplets form in the film and remain
as islands after cooling.  Typical islands have radii from 50 to
several hundred microns.  The boundaries of these inclusions are
smectic dislocations, and the $\cv$ director has a preferred
orientation relative to the tangent line to such dislocations
\cite{Hatwalne95-1} thus providing our fixed boundary condition.  This
fixed boundary condition allows us to consider the interior and
exterior regions of an island as separate systems, analogous to the
way a conducting surface allows one to independently consider the
electrostatic configurations interior and exterior to the conducting
surface.  While island-defect pairs are relatively easy to create and
image (Fig. \ref{setup}), they have the disadvantage that often the
islands shrink, and it is difficult to control their size.  This makes
the creation of several islands of uniform size very difficult.
\par
A second type of inclusion in smectic-C films are isotropic or nematic
droplets, found in materials that have a direct phase transition from
the smectic-C to the isotropic phase.  Droplets have the disadvantage
compared to islands that as one nears the isotropic transition, the
birefringence of the film diminishes, and it becomes difficult to
image the $\cv$-director.  A third type of inclusion consists of the
direct insertion of small (10 micron) glass spheres into the film.
While this in principle is the best technique, since size can be
accurately controlled and the birefringence is unaffected, in practice
it is difficult to maintain glass spheres larger than 5 microns in the
film.  Larger spheres either rupture the film or simply pass through.
\par
Regardless of its type, a circular inclusion with rigid boundary
conditions along its perimeter carries a topological charge of $+1$.
Experimentally it is observed that each island nucleates a companion
$-1$ defect out of the bulk, which maintains a preferred separation
from the inclusion while undergoing visible radial and azimuthal
orientational fluctuations with respect to the center of the
inclusion.
\par
In this paper, we present a theoretical study of these circular
inclusions in smectic-C films.  We also report briefly on preliminary
experimental observations that are in qualitative agreement with our
theoretical analysis.  The theoretical treatment of circular
inclusions with rigid boundary conditions in smectic-C films is
simpler than the treatment of similar defects in Langmuir films and
nematic emulsions since shape changes in the boundary present in
Langmuir films do not occur and within the one coupling constant model
of smectic-C elasticity we avoid the nonlinearities inherent in the
treatment of $3D$ nematics.  In fact, the equation describing the
order is simply Laplace's equation, $\nabla^2 \Phi = 0$, where $\cv =
(\cos\Phi, \sin\Phi)$.  Because we impose rigid boundary conditions,
our $2D$ results are very similar to those obtained in $3D$ nematic
emulsions with similar boundary conditions.  Fluctuations, however,
are more pronounced in $2D$ than in $3D$.
\par
If the angle $\Phi$ is constrained to make a constant angle $\phi_a$
with respect to the tangent to the boundary of an inclusion, then an
inclusion necessarily creates (or ``is'') a topological defect - a
vortex, or equivalently, a disclination with positive topological
charge.  If the boundary conditions at infinity enforce alignment
along a specific direction the total topological charge must be zero,
and, as a consequence, a companion defect with negative charge must be
created for every inclusion.  The inclusion and its companion defect
form a topological dipole that interacts with other distant
inclusion-defect pairs via a $2D$ dipole-dipole interaction.
\par
In $3D$ nematic emulsions, fluctuations in both the distance $r_d$
between a droplet and its companion defect and the angle $\theta$
between the droplet-defect dipole and the direction of the far-field
alignment are negligible, with
\begin{equation}
\left\langle \left( \frac{\delta r_d}{r_d} \right)^2 \right\rangle
\approx \langle (\delta \theta )^2 \rangle \approx \frac{k_B T}{K a} ,
\end{equation}
where $T$ is the temperature, $K$ is a Frank elastic constant and $a$
is the droplet radius ($\sim 1\mu$m). Since $K \approx k_B T/l$,
where $l$ is approximately a molecular length (~$300$nm) we find that
$\langle (\delta \theta )^2 \rangle \approx l/a \approx 0.003
\ll 1.$
\par
In our $2D$ analog, the Frank elastic constant $K_{2D}$ has units of
energy and is typically of order $d K_{3D}$, where $d$ is the film
thickness.  Taking $d=nl$, where $n$ is the number of layers in the
film, we then find $\langle \left( \delta r_d/r_d \right)^2 \rangle
\approx k_B T/K_{2D} \approx 1/n$.  Our calculation actually leads to
the numerically smaller result $\langle \left( \delta r_d/r_d \right)^2 \rangle
\approx 1/8 \pi n$.  In addition, logarithmic singularities in
two-dimensions lead to a logarithmic dependence on the system size,
$R$, in $\langle (\delta \phi)^2 \rangle \approx \log \left(
R/a \right) (k_B T/2 \pi K_{2D})$.  
\par
Laplace's equation in two dimensions admits two fundamental types of
point defects: a topological defect with $\phi = \tan^{-1}(y/x)$ and a
logarithmic defect with $\phi = \log(x^2+y^2)$.  In what follows, we
will refer to the charge of the topological defect as the topological
charge or magnetic charge and that of the logarithmic defect as the
electric charge since the angle field of the latter is equivalent to
the potential produced by a point electric charge (see Appendix).  To
describe arbitrary orientations of the topological dipole with respect
to the $\cv$-director at infinity, both kinds of defects are needed.
The ``charges'' of the topological defects are fixed to be integers
whereas those of the logarithmic defects are proportional to the
angular deviation of the dipole from its favored direction.
\par
As in $3D$ nematic emulsions \cite{Poulin97-1}, widely separated
inclusion-induced topological dipoles interact via a dipole-dipole
potential arising from the arrangement of the magnetic charges
associated with each inclusion.  When the dipoles are not properly
aligned with respect to the $\cv$-director at infinity, the inclusions
will also carry a certain amount of electric charge, leading to an
additional interaction.  In three dimensions, the analogous
interaction is unimportant because of the prohibitive cost of
deviations from the direction of preferred alignment.  In two
dimensions, deviations can be large: this interaction is important,
and it can be calculated.  It predicts that correlated rotations of
dipoles in which the total electric charge is zero are energetically
less costly than those with non-zero electric charge.  It should be
possible to test this prediction experimentally.
\par
Our preliminary experimental observations are in qualitative agreement
with these predictions.  Associated with each island is a companion
negative defect (Fig. \ref{vfs_no_q}).  The position of this defect fluctuates
visibly with angular fluctuations apparently greater than the radial
fluctuations.  Furthermore, defect fluctuations appear to be
correlated when more than one island is present.  Further study will
be needed to provide unambiguous confirmation of these observations.

\section{The Model and Solution}
\par
Our free standing smectic-C film is closely approximated by a $2D$
$XY$-model \cite{Knobler92-1,Pindak82-1}.  A typical film has a small
number of layers, $N$, (roughly 4-10) which can be determined exactly
with laser reflectivity measurements.  The projection of the nematic
director into the plane of the film provides us with our unit-vector
order parameter $\cv$ (the tilt is relatively constant over the area
of the film).  The effective free energy of these films contains two
elastic constants - one for splay distortions ($K_S$) and another for
bend ($K_B$):
\begin{equation}
F = \frac{1}{2} \int_{\Omega} \left[ K_S (\nabla \cdot \cv)^2 + K_B (\nabla
\times \cv)^2 \right] d^2x.
\end{equation}
However, when the difference between $K_B$ and $K_S$ is small, the $2D$
$XY$-model provides an accurate approximation,
\begin{eqnarray}
F &=& \frac{K}{2} \int_{\Omega}  \left[ (\nabla \cdot \cv)^2 + (\nabla \times
\cv)^2 \right] d^2x \\ 
&=& \frac{K}{2} \int_{\Omega} (\nabla \Phi)^2 d^2x,
\label{FreeEnergy}
\end{eqnarray}
where it is understood that the integral is taken over the region
$\Omega$ where the order parameter $\cv$ exists and is nonzero.
\par
Let us now examine an ideal case of a circular inclusion immersed in
an ordered medium described by an $XY$ order parameter $\cv =
(\cos\Phi,\sin\Phi)$.  We take Eq. (\ref{FreeEnergy}) to be our free
energy, where $K$ is the Frank elastic constant and the integral is to
be taken over the region that the ordered medium occupies (external to
the inclusion and internal to some boundary farther away), namely
$\Omega = B_R - B_a$, where $B_r$ is a disc of radius $r$ centered
about the origin, $a$ is the radius of the inclusion and $R$ is the
system size.  We do not include any surface energies since we will
assume the properties of the surface will remain fixed (i.e., strong
pinning boundary conditions keeping $\cv$ fixed and a very large line
tension keeping the shape fixed).
\par
We will assume that the far field is uniform at the outer boundary
taking $\Phi|_{r=R} = 0$.  Next we take the boundary condition at the
surface to be of the form $\Phi|_{r=a}=\phi +\frac{\pi}{2} + \phi_a$,
where $\phi$ is the azimuthal angle relative to the center of the disc
and $\phi_a$ is the angle that the director $\cv$ makes with the
tangent vector to the boundary.  This condition gives the inclusion a
topological charge (winding number) of $+1$.  The far-field boundary
condition requires that the total charge be 0, thus requiring the
existence of an additional topological defect with charge $-1$.  We
will find that this defect will sit fairly close to the inclusion and
that the pair will constitute a dipole, which can in turn interact with
other such pairs via a dipole-dipole interaction.  We will note that
the value of $\phi_a$ will only be relevant if one knows the actual
direction of the far field, and then it will determine the relative
angle between the far-field direction and the direction of the dipole.
\par
Taking the radius of the defect to be $\xi \ll a$ and assigning a core
energy to the defect $\epsilon_{\text{core}} \approx \pi K $
\cite{Chaikin}, we simply need to find solutions for $\Phi$ that depend on
the position of the defect and to extract an effective free energy,
$F[\rv_d]$, as a function of the defect position $\rv_d$. Minimization
of Eq. (\ref{FreeEnergy}) leads to Laplace's equation, $\nabla^2 \Phi
=0$ throughout $\Omega$ except at points occupied by defects.  The
angle field of a single charge $+1$ defect located at $(x_0,y_0) =
(r_0 \cos \phi_0, r_0 \sin \phi_0) = (r_0,\phi_0)=\rv_0$ is simply
\begin{equation}
\phi_{\rv_0}(\rv)=\phi_{\rv_0}(x,y) = {\rm Im}\left(\log|z-z_0|\right),
\label{charge1}
\end{equation}
where $z=x+iy$.  We note that $\phi_{\rv_0}(\rv)$ satisfies Laplace's
equation everywhere except at $\rv_0$.  The angle field of a
charge-$q$ defect at $\rv_0$ is $q \phi_{\rv_0}$.  In particular, a
charge $-1$ defect has a field $-\phi_{\rv_0}$.
\par
There is an image solution to our problem of a disc with strong
pinning boundary conditions and a uniform director field at infinity
\cite{Ranganath76-1,Chandra86-1}:
\begin{equation}
\left( \phi_{\rv_0} + \phi_{\rv_0^{{\rm im}}} \right)
|_{r=a} = \phi +\phi_0 + \pi + 2 k \pi,
\end{equation}
where $\rv_0^{{\rm im}}=(a^2/r_0, \phi_0)$ and $k$ is an integer,
which may vary depending upon the point on the circle $r=a$ at which
we are evaluating the function.  The function
\begin{equation}
\Phi = 2\phi - \phi_{(r_d,\phi_d^0)} - \phi_{(\frac{a^2}{r_d},\phi_d^0)},
\label{Phi_r}  
\end{equation}
where
\begin{equation}
\phi_d^0 = \frac{\pi}{2} - \phi_a,
\nonumber
\end{equation}
has the following properties: $\Phi |_{r=a} = (\phi + \frac{\pi}{2} +
\phi_a) \bmod 2\pi$ and $\Phi \rightarrow 0$ as $r \rightarrow
\infty$.  Thus we can satisfy boundary conditions at $r=a$ and at
$\infty$ by creating a real $-1$ defect at $(r_d,\phi_d^0)$ in the
medium outside the circular inclusion and a $+2$ defect at $r=0$ and a
$-1$ defect at $(a^2/r_d, \phi_d^0)$ inside the disc.  Since in the
absence of boundary conditions at $R$, there would be a single $+1$
defect at $r=0$, we may regard one $+1$ defect at $r=0$ and the
interior $-1$ defect as images.  Even though this solution does not
strictly satisfy our boundary condition at $R$, we claim that it is
good enough for $R$ significantly larger than $a$.  To satisfy both
boundary conditions would require an infinite number of images inside
$a$ and outside $R$, but the additional images would have little
effect on the energy as the interested reader may easily confirm.
\par
\begin{figure}
\psfig{figure=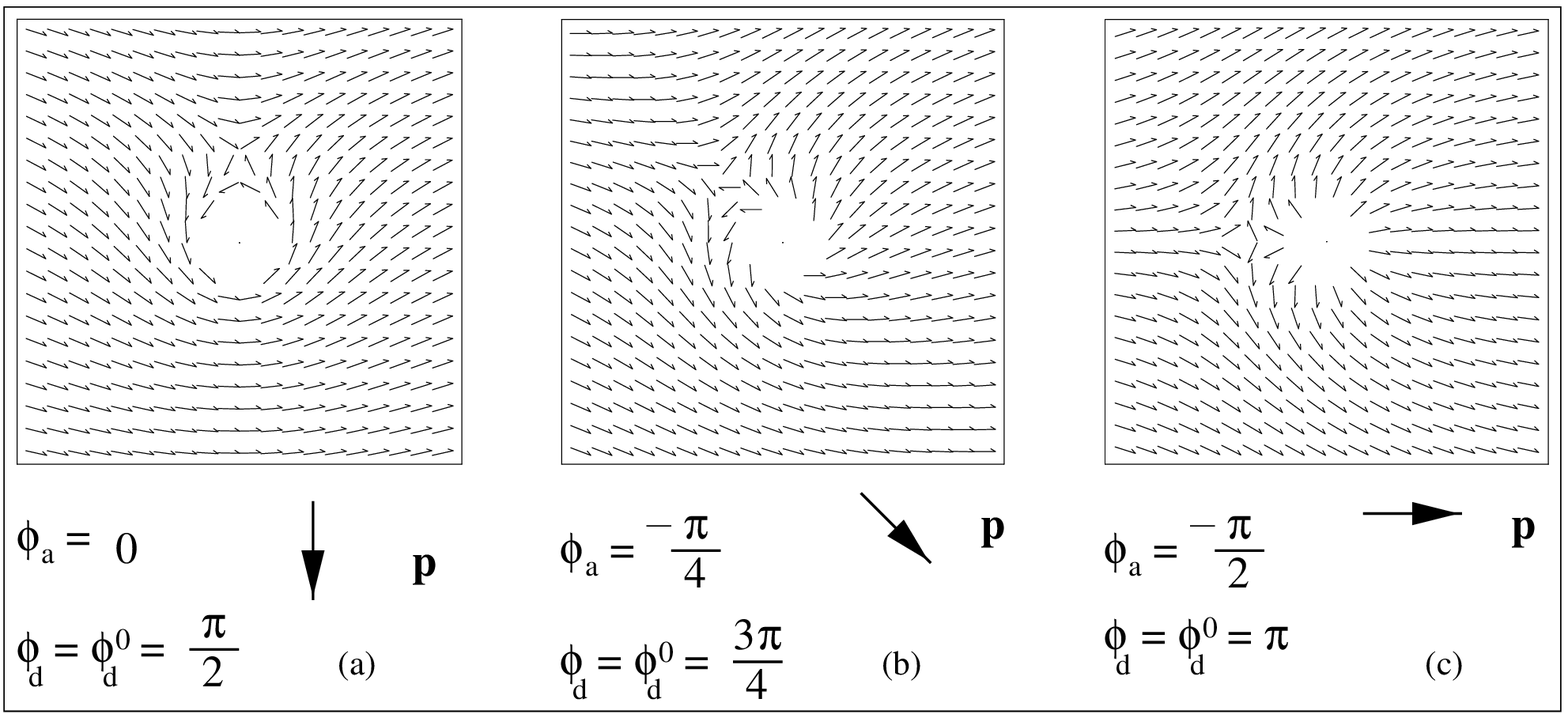,width=16cm}
\caption{
        }
\label{vfs_no_q}
\end{figure}

The collection of three charges can be viewed as a topological dipole
with moment $\pv$ pointing from the induced defect in the medium to
the center of the disc:
\begin{equation}
\pv = -p(\cos\phi_d^0, \sin\phi_d^0), \quad p=r_d+\frac{a^2}{r_d} .
\end{equation}
The dipole makes an angle of $\pi + \phi_d^0$ with the direction of
$\cv$ in the far field.  Sample configurations are shown in Fig
(\ref{vfs_no_q}).  If $\cv$ is parallel to the $x$-axis at infinity,
then the induced defect lies above the disc, and hence $\pv$ points in
the $-y$ direction when $\phi_a=0$ (tangential boundary conditions);
and the induced defect lies to the left of the disc, and hence $\pv$
points in the $+x$ direction when $\phi_a=-\frac{\pi}{2} $ (normal
boundary conditions).  An intermediate configuration with
$\phi_a=-\frac{\pi}{4}$ is also shown in Fig. (\ref{vfs_no_q}).
\par
The solution we have constructed is valid for any $r_d$, and we now
calculate the free energy of a single dipole, $F_1[(a,r_d,\phi_d^0]$,
as a function of $r_d$ at the fixed orientation $\pi + \phi_d^0$
\begin{eqnarray}
\label{F_r}
&&F_1[(a,r_d,\phi_d^0)] = -\pi K \log (a r_d) + 2\pi K\log r_d
\\ \nonumber 
&&\quad -\pi K \log\xi -\pi K \log\left(r_d-\frac{a^2}{r_d}\right) +2
\pi K \log r_d + \epsilon_{\rm core} \\ \nonumber
&&= -\pi K \log \left( \frac{a^2}{r_d^2} -\frac{a^4}{r_d^4} \right) + 
\pi K \log \frac{a}{\xi} + \epsilon_{\rm core}.
\end{eqnarray}
We see that the preferred position for the defect is $r_d^0 =
\sqrt{2} a$.  Expanding about this minimum with $\delta r_d = r_d -
r_d^0$ we find,
\begin{equation}
F[(\delta r_d,\phi_d^0)] \approx {\rm const} + 4 \pi K
\left(\frac{\delta r_d}{r_d^0}\right)^2 + \cdots.
\end{equation}
Thermal fluctuations in $r_d$ then satisfy,
\begin{equation}
\left \langle \left( \frac{\delta r_d}{r_d^0}\right)^2 \right \rangle
\approx \frac{k_B T}{8 \pi K} ,
\label{del_r}
\end{equation}
in qualitative agreement with our estimate from section (I) .  Noting
that this $K$ is the two-dimensional constant $K_{2D}$ from section
(I), we recall the estimate $k_B T/K \approx 1/n$ where $n$ is the
number of layers in the film (typically 4-10).  This would suggest
that the radial fluctuations should be of order $r_d^0/10$ for a $4$
layer film.  Our experimental observations are slightly larger than
this.
\par
We would now like to relax our assumption about the azimuthal location
of the defect and calculate $F[a,\rv_d]$ where $\rv_d=(r_d, \phi_d)$
and $\phi_d = \phi_d^0 + \delta\phi_d$.  Equivalently, we now let the
angle of the dipole ($\pv=-p(\cos\phi_d,\sin\phi_d)$) vary as well as
its magnitude (Fig. \ref{vfs_q}).  In doing so, we will in fact see
that $\phi_d^0$ is the equilibrium azimuthal location of the defect.
To find a solution, we have only to add to $\Phi$ in Eq. (\ref{Phi_r})
a solution to $\nabla^2 \Phi =0$ that is zero at $r=R$ and equal to
$\delta\phi_d$ at $r=a$.  Such a solution can be constructed from the
electric-charge field ${\rm Re}\log z = \log r$, which is constant for
constant $r$.  We find
\begin{equation}
\Phi = 2\phi - \phi_{\rv_d} - \phi_{\rv_d^{{\rm im}}} - q
\log\left(\frac{r}{R}\right),
\end{equation}
where
\begin{equation}
q = \frac{\delta\phi_d}{\log\left(\frac{R}{a}\right)}
\end{equation}
is the electric charge.  The energy associated with this field is
\begin{equation}
F[(a,r_d,\phi_d)] = F_1[(a,r_d,\phi_d^0)] + \frac{\pi K}{\log \frac{R}{a}}
(\delta \phi_d)^2.
\label{F_delphi}
\end{equation}
From this we can calculate the magnitude of the orientational
fluctuations:
\begin{equation}
\langle (\delta \phi_d)^2 \rangle = \frac{k_B T}{2 \pi K} \log \left(
\frac{R}{a} \right).
\label{del_phi}
\end{equation}
Thus, in an infinite sample $\langle (\delta \phi_d)^2 \rangle$ is
infinite, and there is no well-defined direction of $\pv$ relative to
the direction of $\cv$ at infinity.  In laboratory samples one
typically finds $R/a \approx 10^2 - 10^4$ giving $\log (R/a) \approx
5-9$ and hence fluctuations in the angle are reasonably enhanced by
this logarithmic factor.  Finally again using our estimate of $k_B T/K
\approx 1/n$ with $n=4$ we find that for $R/a = 10^3$ we have $\langle (
\delta\phi_d)^2 \rangle \approx 0.25$.  This corresponds to angular
fluctuations of approximately 30 degrees, which should be far more
noticeable than the radial fluctuations $\delta r_d$.  Observed
angular fluctuations indeed appear to be of this order.

\section{Multiple Domains}
\par 
A single inclusion distorts the $\cv$-director field as we have just
shown.  If there are two or more inclusions, the distortions they
induce will give rise to effective interactions among the inclusions.
If the separation between inclusions is large compared to their radii,
each inclusion behaves like a dipole, and there will be a pair-wise
dipole-dipole interaction between them.  In addition, if the dipoles
are not properly oriented with respect to the far-field $\cv$-director
they spawn ``electric'' charges within the inclusion, and these give
rise to a $2D$ Coulomb interaction between the inclusions.  More
importantly, since the values of these electric charges are not fixed,
the usually constant ``self-energy'' contribution will not be constant
and will greatly favor total charge neutrality within the system.
This in turn will imply that fluctuations of the dipoles adhering to
electrostatic charge neutrality will be enhanced over those that do
not.
\par
Before proceeding we wish to clarify our notation for a single
inclusion-defect pair.  For simplicity and without loss of generality,
we will continue to assume that the far-field $\cv$-director points in
the $\ev_x$ direction.  From the previous section we conclude that a
single inclusion-defect pair can be fully described by $[\rv, \pv, a,
\phi_a]$, where $\rv$ is the position of the center of the inclusion
(no longer required to be at the origin), $a$ is the radius of the
inclusion, $\phi_a$ tells us what the boundary condition is at the
inclusion boundary, and
\begin{eqnarray}
&&\pv = -p(\cos\phi_d, \sin\phi_d); \\ &&\phi_d = \phi_d^0 +
\delta\phi_d; \quad \phi_d^0 = \frac{\pi}{2}-\phi_a;
\\
&&p = r_d + \frac{a^2}{r_d}; \quad r_d=r_d^0 + \delta r_d .
\end{eqnarray}
We then found it convenient to code $\delta\phi_d$ via
$q=\delta\phi_d/\log(R/a)$.
\par
Now we would like to calculate the free energy for a system of such
inclusion-defect pairs, $F({[\rv_i, \pv_i, a_i, \phi_{ai}]})$, where we
assume that this description of multiple inclusions is valid when $R
\gg |\rv_i-\rv_j| \gg a$ for all pairs $(i,j)$.  As we are now aware,
each isolated inclusion-defect pair can be thought of purely in terms
of the external topological defect and an arrangement of topological
and logarithmic defects inside the inclusion.  Of course, when there
is more than one inclusion, additional images must be introduced in
the interior of each inclusion to satisfy boundary conditions.  The
full calculation of the energy is then reduced to calculating the
energy from all of the topological charges and electric charges.
As long as the inclusion-defect pairs are reasonably separated,
however, one can obtain a good estimate of the energy by ignoring the
contributions made by the additional image charges.  The energy
decomposes in a natural way into three pieces: the first is the
contribution each individual inclusion-defect pair makes when its
electric charge $q_i=0$, the second is the contribution from the
dipole-dipole interaction among the topological defects, and finally,
the third is the total electrostatic energy arising from the electric
charges:
\begin{eqnarray}
&&F(\{[\rv_i,\pv_i,a_i,\phi_{ai}]\}) \approx \sum_i F_1(a_i,r_{di},\phi_{ai}) \nonumber \\
&& + \sum_{i<j} 2 \pi K \left( \frac{\pv_i \cdot \pv_j}{r_{ij}^2} - \frac{2 (\pv_i
\cdot \rv_{ij}) (\pv_j \cdot \rv_{ij})}{r_{ij}^4} \right) \nonumber \\
&& + \pi K \left(\sum_i q_i\right) \left(\sum_i q_i
\log\left(\frac{R}{a_i}\right) \right) \nonumber \\
&& - \sum_{i<j} \pi K q_i q_j \log \left(\frac{r_{ij}^2}{a_i
a_j}\right),
\label{F_multi}
\end{eqnarray}
where $F_1$ is defined by Eq. (\ref{F_r}) and $\rv_{ij} = \rv_i -
\rv_j$.  The second term is the dipole-dipole energy and the third and
fourth are the electrostatic energy.  Again we emphasize that this is
only an approximate form for the energy valid when the pairs are
sufficiently for apart.
\par
To see the importance of the first electrostatic term in the energy we
now turn to a more specific example.  For simplicity we will assume
that our system consists of only two inclusion-defect pairs: $[0,
\pv_1, a, -\pi/2], [r_s \ev_x, \pv_2, a, -\pi/2]$.  We have taken
the inclusions to have the same radii and to have the same normal
boundary conditions.  In addition we have placed one inclusion at the
origin and the second at $\rv_s$, where $\rv_s$ is parallel to the
preferred orientation of the inclusion-defect dipoles
(Fig. (\ref{fluctuations})).  Assuming that the magnitudes of the
dipoles are fixed at the value for a single inclusion, $r_d^0$, we can
write
\begin{equation}
\pv_i = p (\cos\phi_i, \sin\phi_i) ,
\end{equation}
where $p = a(\sqrt{2} + 1/\sqrt{2})$ and $\phi_i=\pi+\phi_d$ (note
that $\phi_{di}^0=-\pi$ so that $\phi_i^0 = 0$).  This implies each
inclusion-defect pair carries an electric charge
\begin{equation}
q_i = \frac{\phi_i}{ \log\left(\frac{R}{a}\right)}.
\end{equation}
Using Eq. (\ref{F_multi}) we can calculate
\begin{eqnarray}
F[q_1,q_2,\rv_s] &=& 2 F_1[(a,r_d,\frac{\pi}{2} - \phi_a)] \\ \nonumber
&+& 2 \pi K \left( \frac{\pv_1 \cdot \pv_2}{r_s^2} - \frac{2 (\pv_1
\cdot \rv_s) (\pv_2 \cdot \rv_s)}{r_s^4} \right), \\ \nonumber
&+& \pi K (q_1+q_2)^2 \log\left(\frac{R}{a}\right) - 2 \pi K q_1 q_2
\log \left(\frac{r_s}{a}\right) ,
\end{eqnarray}
where again the first term is simply twice the energy from
Eq.(\ref{F_r}) for a single inclusion with $\delta \phi_d =0$.
Taking $q_{+} = (q_1+q_2)/2$ and $q_{-} = (q_1-q_2)/2 =
q_1-q_{+}$,  we can simplify now to find
\begin{eqnarray}
F[q_{+},q_{-},\rv_s] &=& 2 F_1[(a,r_d,\frac{\pi}{2} - \phi_a)] \\ \nonumber
&+& - 2 \pi K \frac{p^2}{r_s^2} \cos\left(2 q_{+}
\log\left(\frac{R}{a}\right) \right) \\ \nonumber
&+& 2 \pi K q_{+}^2 \left( \log \left(\frac{R}{a} \right) + \log
\left(\frac{R}{r_s} \right) \right) \\ \nonumber
&+& 2 \pi K q_{-}^2 \log \left(\frac{r_s}{a} \right).
\end{eqnarray}
We see immediately that $q_{+}=q_{-}=0$ is the minimum, which
corresponds to $\phi_1=\phi_2=0$ where the dipoles are aligned with
the $+x$-axis.  More interestingly we see that $q_{+}$ and $q_{-}$ are
decoupled.  The equipartition theorem then implies
\begin{eqnarray}
\langle q_{+}^2\rangle &\approx& \frac{k_B T}{4\pi K \left(
\frac{p^2}{r_s^2} \left( \log\left(\frac{R}{a}\right) \right)^2 + 
\log\left(\frac{R}{a}\right) + \log\left(\frac{R}{r_s}\right) \right) } \\ 
\langle q_{-}^2 \rangle &=& \frac{k_B T}{4 \pi K
\log\left(\frac{r_s}{a}\right)}. 
\end{eqnarray}
For the $\phi_i$'s we have,
\begin{eqnarray}
\langle \delta\phi_1^2 \rangle &=& \langle \delta\phi_2^2 \rangle = \left( \langle
q_{+}^2 \rangle + \langle q_{-}^2 \rangle \right)
\left(\log\left(\frac{R}{a}\right)\right)^2 \\   
\langle \delta\phi_1 \delta\phi_2 \rangle &=&  \left( \langle q_{+}^2 \rangle -
\langle q_{-}^2 \rangle \right)
\left(\log\left(\frac{R}{a}\right)\right)^2. 
\end{eqnarray}
In the limit $R \rightarrow \infty$, fluctuations in $q_{+}$ tend to
zero.  The ratio of $\delta\phi_i$ to $q_i$, however, diverges so that
in this limit,
\begin{equation}
\langle (\delta\phi_1 + \delta\phi_2)^2 \rangle = \frac{k_B T}{4 \pi
K}
\end{equation}
is finite and
\begin{equation}
\langle (\delta\phi_1-\delta\phi_2)^2 \rangle = \frac{k_B T}{4 \pi K}
\frac{\log(R/a)}{\log(r_s/a)}
\end{equation}
diverges.
Note that fluctuations in $\langle (\delta\phi_1 - \delta\phi_2)^2
\rangle$ in the two inclusion system are much larger than fluctuations
in $\delta\phi_d$ for a single inclusion.  Of course, the ratio $R/a$
in a real system is such that $\log(R/a)$ is not much greater than
$1$.  Let us compare $\langle \delta\phi_i^2 \rangle$ and $\langle
\delta\phi_1 \delta\phi_2 \rangle$ to $\langle (\delta
\phi_d)^2 \rangle$ from equation (\ref{del_phi}) for realistic
values $R/a = 10^3$ and $r_s/a = 10$.  We find that $\langle
\delta\phi_i^2 \rangle \approx (1.75) \langle (\delta\phi_d)^2
\rangle$ and that $\langle \delta\phi_1 \delta\phi_2 \rangle \approx
-(1.25) \langle (\delta\phi_d)^2 \rangle$.  For $n=4$ this would
correspond to angular fluctuations of a single dipole of approximately
40 degrees versus angular fluctuations of only 30 degrees for an
isolated domain. Furthermore we see that the fluctuations are highly
correlated.  Thus we see that the presence of multiple domains can
enhance the magnitude of the fluctuations a noticeable degree.
\par
Had we not taken the separation vector to be in the direction of the
far-field director we would have first been obliged to find the new
equilibrium values for $q_{+}$ and $q_{-}$.  Not surprisingly we would
have found that the $q_{+}$ equilibrium value would no longer be zero,
however the term in the energy for $q_{-}$ would remain the same.
Since the fluctuations in $q_{-}$ dominate the fluctuations in $q_{+}$
we would still expect the above observations concerning the magnitude
of the fluctuations to be fairly accurate.  Furthermore, in the
presence of many domains, the correlations between them may be quite
complicated but we will still expect some enhancement in the angular
fluctuations arising from those fluctuations which preserve
electric charge neutrality.

\section{Summary and Conclusions}
\par
Smectic-C films are adequately described by the $2D$-$XY$ model.
Circular inclusions in the film with appropriate boundary conditions
carry a topological charge.  Global boundary conditions require the
inclusions to nucleate companion topological defects out of the
surrounding ordered medium, providing a good method for studying the
properties of a small collection of defects.  Herein we have described
how director configurations can be described systematically by
externally nucleated topological defects and associated image defects
lying within the inclusions, provided we allow for logarithmic as well
as topological defects.  The electric charge required is determined
by the orientation of the topological dipole created by the inclusion
and its companion and by the direction of the far-field director.  The
value of the electric charge fluctuates producing visible
fluctuations of the position of the companion defect about its host
inclusion.  When multiple inclusions are present they interact through
their topological and electric charge distributions.  The
topological dipole distributions associated with each inclusion-defect
pair lead to a dipole-dipole interaction between them an to an
alignment of the dipoles.  However, thermal fluctuations of the dipole
direction are large (the fluctuations the companion about the host)
and carry electric charges with them.  Orientational fluctuations
of the dipoles are coupled via the interaction of these electric
charges.  By far the strongest correlations arise from the overriding
desire for global electric charge neutrality in the system (the
energy of an electric charge in $2D$ diverges with the system size).
We have presented detailed calculations describing this effect for the
case of two inclusions, and have presented how to approach the problem
for additional inclusions.
\par
We have considered here only the case of zero total topological charge
brought about by parallel boundary conditions at infinity.  If the
$\cv$-director has some fixed angle relative to the outer boundaries
of the sample, then the total topological charge is $+1$.  In this
case, a single inclusion will not force the nucleation of a companion
defect.  Each additional inclusion will however, and there will be
$(n-1)$ defects in a sample with $n$ inclusions.  This is entirely
analogous to the situation in multiple nematic emulsions where water
droplets are captured inside spherical nematic drops

\acknowledgements
T.C. Lubensky and Darren Link thank Noel Clark for helpful
discussions. The authors were supported primarily by the Materials
Research Science and Engineering Center Program of NSF under award
number DMR96-32598.  Darren Link acknowledges further support from
NASA under NAG3-1846 as well as support from the NSF under DMR96-14061.

\newpage
\appendix
\section{An Electromagnetic Analogy}
\par
We take the most general $\Phi$ to be of the form
\begin{equation}
\Phi = \sum_{i} \left(q_i^E \log|\rv-\rv_i| + q_i^B \phi_{\rv_i}(\rv)
\right).
\end{equation}
It can be shown that
\begin{equation}
\int (\nabla \log|\rv-\rv_i|)\cdot(\nabla \phi_{\rv_j}(\rv)) d^2x =0
\end{equation}
and thus if we define
\begin{equation}
\Phi^E = \sum_i q_i^E \log|\rv-\rv_i| \qquad \Phi^B = \sum_i q_i^B
\phi_{\rv_i}(\rv) 
\end{equation}
then we have
\begin{equation}
\int (\nabla\Phi)^2 d^2x = \int \left((\nabla\Phi^E)^2
+(\nabla\Phi^B)^2\right) d^2x
\end{equation}  
and we see that the two families of singularities do not interact.
\par
Now recall that the potential for a line charge parallel to the
$z-$axis is
\begin{equation}
\lambda \log\left(\sqrt{(x-x_0)^2+(y-y_0)^2}\right)
\end{equation}
where $(x_0,y_0)$ are the $(x,y)$ coordinates where the line pierces
the $xy-$plane.  Thus, we can identify $\Phi^E$ with a collection of
line charges or equivalently $2D$ electrostatic charges.  Next, noting
that for any function $f(x,y)$ we have
\begin{equation}
\nabla f \cdot \nabla f = (\nabla \times f\ev_z) \cdot (\nabla
\times f\ev_z)
\end{equation}
we can define $\Av=\Phi^B \ev_z$ and note that
\begin{equation}
\left(\nabla \times \Av\right)^2 = \left(\nabla \Phi^B\right)^2.
\end{equation}
Interestingly, the vector potential for a line of magnetic charges
parallel to the $z-$axis piercing the $xy-$plane at $(x_0,y_0)$ is
simply
\begin{equation}
\Av = \tan^{-1}\left( \frac{y-y_0}{x-x_0} \right) \ev_z
\end{equation}
which corresponds precisely to the singularities in $\Phi^B$.
Equivalently we can view this as a magnetic charge in $2D$, just as we
viewed the electrostatic line charge as a charge in $2D$
electrostatics.  Now, making the final identification, 
\begin{equation}
\Ev = -\nabla \Phi^E \qquad \Bv = \nabla \times \Av
\end{equation}
we can write 
\begin{equation}
F = \frac{K}{2} \int \left( \Ev^2 + \Bv^2 \right) d^2x
\end{equation}
which looks just like the energy density for electromagnetism.
Furthermore, we can define electric and magnetic charge densities
\begin{equation}
\rho^E = \sum_i q_i^E \delta(\rv-\rv_i) \qquad \rho^B = \sum_i q_i^B
\delta(\rv-\rv_i)
\end{equation}
and then complete the electromagnetic analogy by writing
\begin{eqnarray}
\Ev = - \nabla \Phi^E & \qquad &\Bv = \nabla \times \Av \\
\nabla \cdot \Ev = 2\pi \rho^E & \qquad & \nabla \cdot \Bv = 2\pi\rho^B.
\end{eqnarray}
This is why we have chosen to call the logarithmic singularities
``electric charges'' and the $\tan^{-1}$ singularities ``magnetic
charges''.  Of course, in this analogy we have no reason to expect the
lines of magnetic charge to be quantized, whereas in the original
problem there is a clear explanation as to why the $\tan^{-1}$
singularities are quantized (the requirement that the physical field
$\Phi$ be single-valued).
\par
Alternatively we could have exploited the fact that
\begin{equation}
(\nabla \phi_{\rv_0}(\rv))^2 = (\nabla \log|\rv-\rv_0|)^2,
\end{equation}
treated the $\tan^{-1}$ singularities as a separate species
of logarithmic singularities and viewed our free energy as being the
sum of two distinct non-interacting electrostatic energies.

\bibliographystyle{prsty}

\begin{figure}
\psfig{figure=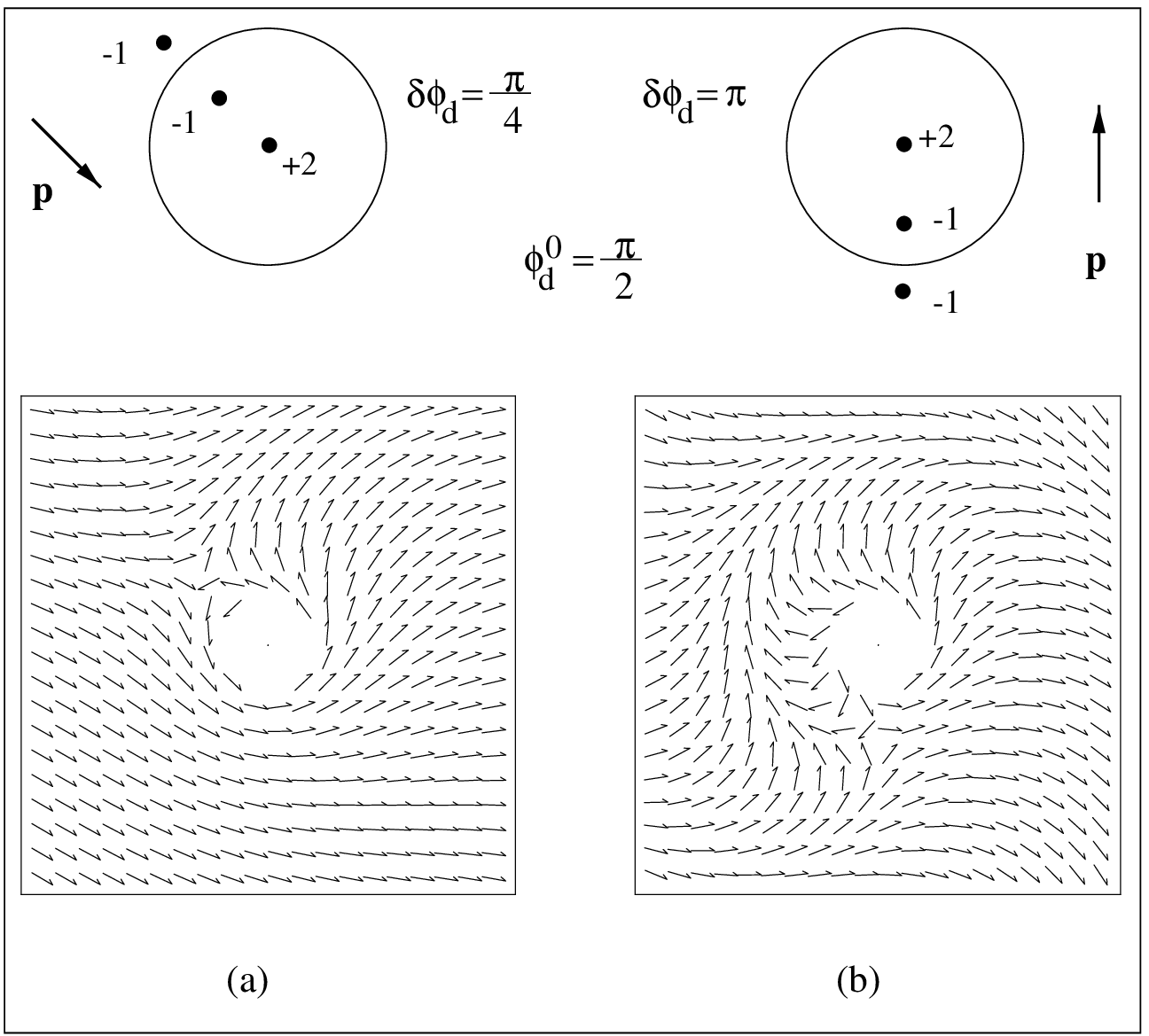,width=8.1cm}
\caption{
        }
\label{vfs_q}
\end{figure}
\begin{figure}
\psfig{figure=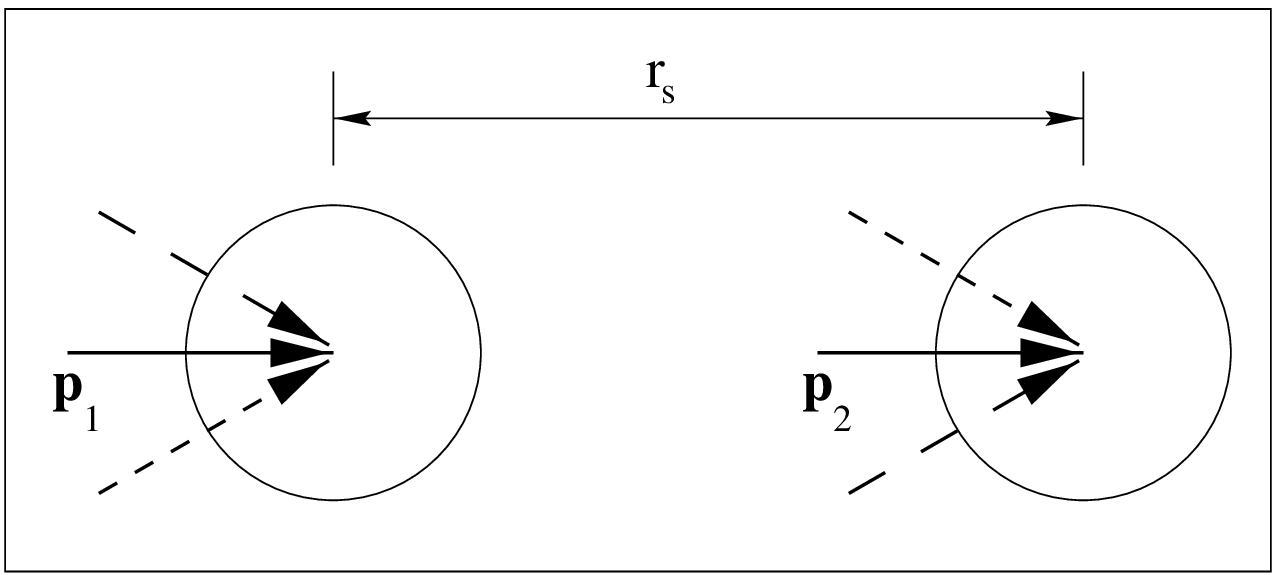,width=8.1cm}
\caption{ 
        }
\label{fluctuations}
\end{figure}

\newpage
Caption Fig.1 \\
\\
(a) A schematic overhead view of a generic circular inclusion, showing
the relative positions of the physical and virtual defects, as well as
the relationships between the far-field director, the director at the
boundary of the inclusion and the direction of the topological dipole
moment.  The director $\cv$ is shown at the boundary and its angle
with respect to the tangent vector $\tv$ measured in a
counterclockwise fashion is $\phi_a$.  The direction of the dipole
moment $\pv$ is given by the usual convention (i.e. from negative to
positive charges).  We see from this and figure (2) that $\pv$ makes
an angle of $-\frac{\pi}{2} - \phi_a$ with respect to the far-field
director. (b) An oblique view of a schematic representation of a film
containing an inclusion consisting of one additional layer.  The rods
represent the actual nematogens in the film which are tilted with
respect to the normal to the smectic planes.  On the top plane the
projection of the nematogens into the smectic plane is represented
using the conventional nails with heads representation, where the head
of the nail distinguishes which end of the projection into the plane
corresponds to the end touching the plane.  The companion defect is
centered at the dark circle.  The figure to the right indicates the
two dimensional interpretation, consisting of the boundary of the
circular inclusion, the direction of the far-field director as well as
the orientation of the director at the boundary of the inclusion, the
positions of the physical and virtual defects and the resulting
topological dipole moment.  (c) Photograph of a four layer freely
suspended smectic-C liquid crystal film with a circular five layer
island of radius 75 microns.  The companion $-1$ defect is seen
sitting outside the island.  The collection of image charges is shown
in the inset.  The presence of only a single $+1$ point defect in the
center of the island in the photograph reminds us that the charges
that the companion $-1$ defect interacts with are image charges
imposed by the boundary conditions and not real physical charges
inside the island.  As mentioned in the paper, the rigid boundary
conditions along the dislocation separating the bulk from the interior
of the island allows us to consider the regions independently.  The
visible $-1$ defect has been observed undergoing large fluctuations in
its radial position as well as its azimuthal orientation.
\\
\\
Caption Fig.2 \\
\\
Three examples of the director field $\cv$ around an inclusion for
three different values of $\phi_a$.  The direction of $\pv$ is
indicated.  In each case we have placed the external defect at its
preferred position, in particular $\delta\phi_d=0$ and hence
$\phi_d=\phi_d^0$.  The values of $\phi_d$, the azimuthal position of
the companion defect, are indicated.  Fig. \ref{vfs_q} shows examples
of fields where $\delta\phi_d$ is nonzero.  Also note that in each
picture the far-field director is taken to be in the $+x$ direction.
\\
\\
Caption Fig.3 \\
\\
Director configurations with nonzero values of $\delta\phi_d$.  The
far-field director is assumed to be in the positive $x$ direction, and
$\phi_a=0$.  Thus the equilibrium position of the companion defect is
above the inclusion ($\phi_d^0=\pi/2$) and the equilibrium direction
of $\pv$ is down.  In (a) $\delta\phi_d=\pi/4$ and in (b)
$\delta\phi_d=\pi$.  Compared to the equilibrium configuration of
Fig. 2(a), note the increasing distortion of the $\cv$-director as
$\delta\phi_d$ increases.  The energy of (b) with $\pv$ pointing up is
clearly higher than that with $\pv$ down.  Unlike $3D$ nematics, there
is a preferred direction rather than a preferred axis of orientation
for $\pv$.  This is not surprising since a nematic is invariant under
$\nv \rightarrow -\nv$ whereas a smectic-C is not invariant under $\cv
\rightarrow -\cv$.
\\
\\
Caption Fig.4 \\
\\
 Above we have two interacting inclusions.  The equilibrium
configuration for the dipoles is represented by the solid arrows.  The
dashed arrows represent possible orientations of the dipoles due to
thermal fluctuations.  Note the correlations in the directions of the
dipole moments, when $\pv_1$ is tilted downward we expect to find
$\pv_2$ tilted upward.

\onecolumn
\newpage

\newpage

\newpage

\newpage

\end{document}